\documentclass{article}
\usepackage{frascatiphys}
\begin{document}

$ $

\indent

\begin{flushright}
CPT-2004/P.043
\end{flushright}

\indent

\indent

\indent

\begin{center}
{\bf{\large{CHIRAL PERTURBATION THEORY \\
\vspace{2 mm} 
CONFRONTED WITH EXPERIMENT}}}

\indent

{\bf Marc Knecht} 

\indent
 
{\em Centre de Physique Th\'eorique}
\footnote{
Unit\'e Mixte de Recherche (UMR 6207) du CNRS, et des 
universit\'es Aix-Marseille I, Aix-Marseille II et du Sud Toulon-Var.
Laboratoire affili\'e \`a la FRUMAM (FR 2291).} \\
{\em CNRS Luminy, Case 907 \\
F-13288 Marseille cedex 9}

\indent

\indent

\indent

\indent

\bf{Abstract}

\end{center}

\noindent
The general framework and the present status of the low energy
theory of the standard model are briefly reviewed. 
Recent applications to a few topics of interest 
for the determinations of 
$\vert V_{ud}\vert$ and of $\vert V_{us}\vert$ are discussed.

\indent

\indent

\indent

\indent

\indent

\indent

\indent

\indent

\indent

\noindent
{\em Talk given at the DA$\Phi$NE 2004 workshop, Frascati, June 7 - 11, 2004}

\hfill
\newpage
\title{CHIRAL PERTURBATION THEORY CONFRONTED WITH EXPERIMENT
}
\author{
Marc Knecht \\
{\em Centre de Physique Th\'eorique,
%\footnote{
%Unit\'e Mixte de Recherche (UMR 6207) du CNRS, et des 
%universit\'es Aix-Marseille I, Aix-Marseille II et du Sud Toulon-Var.
%Laboratoire affili\'e \`a la FRUMAM (FR 2291).} 
CNRS Luminy, Case 907, 
F-13288 Marseille}
}
\maketitle
\baselineskip=11.6pt
\begin{abstract}
The general framework and the present status of the low energy
theory of the standard model are briefly reviewed. 
Recent applications to a few topics of interest 
for the determinations of 
$\vert V_{ud}\vert$ and of $\vert V_{us}\vert$ are discussed.
\end{abstract}
\baselineskip=14pt
\section{Low energy theory of the standard model}
At low energies, the standard model can be described
in terms of an effective theory, involving only the
lightest states as explicit degrees of freedom.
In order that such an effective description becomes possible,
two requirements need to be met. First, one must
have a clear separation of scales (mass gap) between, 
on the one side, the light states, 
%which will appear as 
%explicit degrees of freedom in the effective theory, 
and, on the other side, the heavy states, which appear 
only indirectly in the effective theory,
through their contribution to the infinite number of 
couplings, the low energy constants (LECs) describing the local 
interactions of the light states. The second requirement
is that the masses of the light degrees of freedom are
protected by some symmetry, in order that their lightness
appears as natural, in the very precise sense defined by 
't~Hooft\cite{'tHooft:1979bh} some time ago. In practice,
this means that light spin 0 states have to correspond to
Goldstone bosons produced by the spontaneous breaking
of some continuous global symmetry. The masses of light
fermion will be protected by chiral symmetry, whereas
gauge invariance will ensure that spin 1 gauge fields
remain massless (or massive but light, in the presence of
a Higgs mechanism). 

In the case of the standard model, the light degrees of 
freedom that can be identified in this way comprise:
i) the pseudoscalar meson octet, $\pi$, $K$ and $\eta$,
which, in the limit of massless quarks, become the
Goldstone bosons associated with the spontaneous breaking 
of the chiral symmetry of QCD, ii) the light leptons, 
$e^{\pm}$, $\mu^{\pm}$ and their neutrinos (in 
principle, one might add the $\tau$ neutrino to this list, 
although the $\tau$ lepton itself belongs to the heavy 
states in the context of the present discussion), and
iii) the photon. The range of applicability of this 
effective theory is limited by the typical mass scale 
$\Lambda_H\sim 1$ GeV provided by the non Goldstone mesonic 
bound states. Notice that according to the criteria adopted
above, other effective theories could be considered, for 
instance the one involving only the electron, the three 
neutrinos, and the photon, with the limiting mass scale 
set by $m_\mu\sim M_\pi$, etc.

Chiral perturbation theory\cite{Weinberg:1978kz,GL84,GL85} 
(ChPT) organizes the low energy effective 
theory in a systematic expansion in powers of momenta and of
light
\begin{table}[h]
\centering
\caption{ \it The low energy constants corresponding to some of the parts
of ${\cal L}_{\mbox{\scriptsize{eff}}}$ that have been constructed. They
allow for a description of meson scattering amplitude and meson form
factors up to two loops, anf for the inclusion of ${\cal O}(\alpha)$ radiative
corrections up to one loop.
}
\vskip 0.1 in
\begin{tabular}{|l|c|c|} \hline
          &  2 flavours & 3 flavours \\
\hline
\hline
 ${\cal O}(p^2)$   
    & $F$, $B$  
    & $F_0$, $B_0$   \\
 ${\cal O}(p^4)$   
    & $h_1$,$h_2$, $h_3$, $l_i$, $i=1\dots 7$ \protect\cite{GL84}
    &  $H_1$,$H_2$, $L_i$, $i=1\dots 10$ \protect\cite{GL85}  \\ 
 ${\cal O}(p^6)$   
    & $c_i$, $i=1\dots 57$ \protect\cite{Bijnens99}
    & $C_i$, $i=1\dots 94$ \protect\cite{Bijnens99}   \\
\hline
${\cal O}(\alpha p^0)$   
    & $Z$ \protect\cite{Ecker:1988te,KnechtUrech}  
    & $Z$ \protect\cite{Ecker:1988te}   \\  
 ${\cal O}(\alpha p^2)$   
    & $k_i$, $i=1\dots 11$ \protect\cite{KnechtUrech}
    & $K_i$, $i=1\dots 14$\protect\cite{Urech95}, 
                     $X_i$, $i=1\dots 8$\protect\cite{Knecht:1999ag}   \\
\hline
\end{tabular}
\label{tab1}
\end{table}
masses. The most convenient tool to materialize this expansion 
is to construct an effective lagrangian $
%\begin{equation}
{\cal L}_{\mbox{\scriptsize{eff}}} = {\cal L}_2 
+ {\cal L}_4 + \cdots
%\label{Leff} 
%\end{equation}
$,
where ${\cal L}_n$ contains all the terms of order
$\delta^n$, with $\delta \sim p/\Lambda_H \sim 
M_P/\Lambda_H \sim m_{\ell}/\Lambda_H \sim e$, 
for instance, modulated by LECs whose values depend
on the dynamical properties of the heavy degrees of
freedom that have been integrated out. At lowest order, 
one only needs to compute tree graphs generated by
${\cal L}_2$, whereas the NLO involves both tree graphs
from ${\cal L}_4$ and one loop graphs, and so on. It is
essential to include the loop graphs, with increasing 
number of loops at each new order, in order to correctly 
account for all the singularities (poles, cuts) coming from the light 
degrees of freedom. Computing higher orders in the effective
theory potentially increases the theoretical precision.
However, the number of LECs also increases, as shown
in Table \ref{tab1}. Predictions can thus only be made if some 
knowledge about their values is available. How this problem
can be adressed in practice will be illustrated in the case
of the few examples discussed below.
% 
%
%%%%%%%%%%%%%%%%%%%%%%%%%%%%
\section{Radiative corrections to $\pi_{\ell 2}$, 
$K_{\ell 2}$, and $K_{\ell 3}$ decay modes}
%%%%%%%%%%%%%%%%%%%%%%%%%%%%
As a first application, let us consider the ${\cal O}(\alpha)$
electromagnetic contributions to the semileptonic decays of the 
pion and the kaon.
The general structure of the $\pi_{\ell 2}$ and $K_{\ell 2}$ 
decay rates with radiative corrections included is 
known\cite{Sirlin93}
%\begin{equation}
%\Gamma_{\pi_{\ell 2}(\gamma)} = 
%\frac{G_{\mu}^2}{8\pi}\vert V_{ud}\vert^2 F_{\pi}^2 
%m_\ell^2 \bigg(1 - \frac{m_\ell^2}{M_\pi^2}\bigg)^2
%\times
%\bigg[ 1 + \frac{\alpha}{\pi} C_\pi + {\cal O}(\alpha^2)
%\bigg]
%\end{equation}
%
\begin{equation}
\Gamma_{P_{\ell 2}(\gamma)} = 
\frac{G_{\mu}^2}{8\pi}\vert V_{CKM}\vert^2 F_{P}^2 
m_\ell^2 \bigg(1 - \frac{m_\ell^2}{M_P^2}\bigg)^2
\times
\bigg[ 1 + \frac{\alpha}{\pi} C_P + {\cal O}(\alpha^2)
\bigg]
\end{equation}
with $(P,V_{CKM}) = (\pi, V_{ud})$ or $(K,V_{us})$.
ChPT reproduces this structure, with $C_{P} =
C_{P}^{(0)} + C_{P}^{(2)} + \dots$. The
expressions\cite{Neufeld:1995mu,Knecht:1999ag} for the ${\cal O}(p^0)$ 
contributions $C_{\pi,K}^{(0)}$ involve a (common)
short distance logarithm\cite{Sirlin93},
chiral logarithms, and the low energy constants $K_i$ 
and $X_i$, while $C_{\pi,K}^{(2)}$ and higher represent 
SU(3) breaking quark mass corrections. Interestingly, 
the contributions of the low energy constants drop 
out\cite{Knecht:1999ag} in the ${\cal O}(\alpha)$ correction 
to $\Gamma_{K_{\ell 2}(\gamma)}/\Gamma_{\pi_{\ell 2}(\gamma)}$,
\begin{equation}
C_{\pi} - C_K = \frac{Z}{4} \ln \frac{M_K^2}{M_\pi^2}
 +  {\cal O}(M_K^2/\Lambda_H^2) 
= 0.50 \pm 0.15\,,
\label{CKCpi}
\end{equation}
with $Z$ given by $M_{\pi^\pm}^2 - M_{\pi^0}^2 = 
2e^2 F_\pi^2 Z$, and the error is a conservative 
estimate for SU(3) breaking corrections. This then
leads to 
\begin{equation}
\left\vert\frac{V_{us}}{V_{ud}}\right\vert^2\,\frac{F_K^2}{F_\pi^2}
\,=\,( 7558 \pm 23 \pm 3)\times 10^{-5}\,,
\label{FKFpi}
\end{equation}
where the first error comes from the experimental uncertainties
on the decay rates, and the second error comes from
Eq. (\ref{CKCpi}).

Turning now to $K_{\ell 3}$, the general structure of the amplitudes reads
\begin{equation}
{\cal M}^{(0)}(K_{\ell 3})=G_\mu V_{us}^* C_{CG}L^{\mu}
\big[
f_+(t)(p_K + p_\pi)_\mu + f_-(t)(p_K - p_\pi)_\mu
\big]\,.
\label{Kl3amp}
\end{equation}
For the $K_{e 3}$ modes, only $f_+(t)$ needs
to be considered, whereas for the $K_{\mu 3}$ modes
$f_-(t)$ has to be included as well. The chiral expansions
of these form factors read $f_+ = 1 + f_+^{(2)} +  f_+^{(4)}
+\dots$
and $f_- = f_-^{(2)} + f_-^{(4)} + \dots$ The one loop corrections
$f_\pm^{(2)}(t)$ arising from mesonic intermediate states, including
isospin breaking effects induced by $m_u\neq m_d$, 
are known\cite{Gasser:1984ux,LeutwylerRoos} for quite some time. 
Including ${\cal O}(\alpha)$ radiative 
corrections\cite{Cirigliano:2001mk,Cirigliano:2004pv}
amounts to replacing $f_{\pm}(t)$ by
\begin{equation}
F_{\pm}(t,v) = \big[1 + \frac{\alpha}{\pi}
\Gamma(v,m_\gamma)\big]\times
\big({\widetilde f}_{\pm}(t) +
{\widehat f}_{\pm}(t)\big)\,.
\end{equation}
In this expression, ${\widetilde f}_{\pm}(t)$
contains corrections from the loops and from
$\pi^0 - \eta$ mixing, while ${\widehat f}_{\pm}(t)$ 
collects the remaining counterterm contributions.
Finally, $\Gamma(v,m_\gamma)$, with $v=(p_K - p_\pi)^2$
for $K_{\ell 3}^\pm$, and $v=(p_K - p_\pi)^2$ for 
$K_{\ell 3}^0$, contains the long distance components
of the loops with a virtual photon. The IR divergence,
materialized by the dependence on the photon mass 
$m_\gamma$, is cancelled upon considering the 
differential rates with the emission of a real
soft photon. Corrections at order ${\cal O}(\alpha p^2)$ were
computed\cite{Cirigliano:2001mk,Cirigliano:2004pv} and
the corresponding numerical estimates read
\begin{equation}
{\widetilde f}_{\pm}(0) = 1.0002\pm 0.0022
\,,\ {\widehat f}_{\pm}(0) = 0.0032\pm 0.0016
\quad [K^\pm]
\end{equation}
\begin{equation}
{\widetilde f}_{\pm}(0) = 0.097699 \pm 0.00002
\,,\ {\widehat f}_{\pm}(0) = 0.0046\pm 0.0008
\quad [K^0]
\end{equation}
The expressions of the two loop corrections $f_\pm^{(4)}(t)$
were worked out\cite{Bijnens03} in the isospin limit, and will be 
discussed below. 
% 
%
%%%%%%%%%%%%%%%
\section{The pion beta decay $\pi^+\to\pi^0 e^+\nu_e$ and $\vert V_{ud}\vert$}
%%%%%%%%%%%%%%%
The beta decay of the charged pion ($\pi\beta$) in principle provides 
a determination of $\vert V_{ud}\vert$ which combines
the advantages of the superallowed nuclear Fermi transitions
(pure vector transition, no axial vector admixture), and  of
the neutron beta decay (no nuclear structure dependent radiative 
corrections). There is however a serious drawback, the tiny
branching ratio, Br$(\pi\beta)\sim 1\times 10^{-8}$.
In the absence of radiative corrections, the amplitude has the
structure given in Eq. (\ref{Kl3amp}), with $V_{us}$ replaced
by $V_{ud}$, and $f_\pm(t)$ replaced by  $f^{\pi\beta}_\pm(t)$. 
Contribution from $f^{\pi\beta}_-(t)$ are suppressed 
by $m_e^2/M_\pi^2$ and can be neglected. Furthermore, 
$f^{\pi\beta}_+(t) = 1 + f_{\pi\beta}^{(2)}(t) + \dots$, where the one loop 
corrections\cite{Cirigliano:2002ng} to the CVC result are
small, $f_{\pi\beta}^{(2)}(0) = -7\times 10^{-6}$.
%, since
%they must be quadratic to the isospin breaking pion
%mass difference\cite{BehrendsSirlin,AdemolloGatto}. 
As a consequence,
higher order corrections, $f_{\pi\beta}^{(4)}(0)$, etc., can
be safely neglected. On the other hand, radiative corrections
then become relevant. Including ${\cal O}(\alpha p^2)$
effects gives\cite{Cirigliano:2002ng}
\begin{equation}
\vert V_{ud}\vert \cdot \vert f_+^{\pi\beta}(0)\vert=
9600.8\sqrt{{\mbox{Br}}(\pi^+\to\pi^0 e^+\nu_e(\gamma))}
%\big/
%\vert f_+^{\pi\beta}(0)\vert 
,\,
f_+^{\pi\beta}(0) = 1.0046\pm 0.0005.
\end{equation}
%and $\lambda_+^{\pi\beta} = 0.037\pm 0.003\,.
Radiative corrections enhance the branching 
ratio by $(3.34\pm 0.10)\%$. The (very small)
uncertainties come from the counterterm contributions. 
It is thus possible to give a very accurate prediction 
for $\vert f_+^{\pi\beta}(0)\vert$ in ChPT. With the latest
result\cite{PIBETA04} of the PIBETA experiment, the relative
precision on $\vert V_{ud}\vert$ obtained this way
is still limited by the experimental precision
\begin{equation}
\delta\vert V_{ud}\vert/\vert V_{ud}\vert
= (\pm 3.2_{\,\mbox{\scriptsize{exp}}}
\pm 0.5_{\,\mbox{\scriptsize{th}}})\times 10^{-3}\,.
\end{equation}
% 
%
%%%%%%%%%%%%%%%%%%%%%%%%%%%%%%%%%%%%%%
\section{Two loop $K_{\ell 3}$ form factors and strategies to 
extract $\vert V_{us}\vert$}
%%%%%%%%%%%%%%%%%%%%%%%%%%%%%%%%%%%%%%
The situation is somewhat less ideal for the $K_{\ell 3}$ decays,
since the corrections are larger, and the one loop result is not sufficient
for an accurate determination\cite{LeutwylerRoos} of $\vert V_{us}\vert$. 
The NNLO expressions for the $K_{\ell 3}$ form factors $f_\pm(t)$
decompose into a two loop part, which depends only on the masses and on
$F_\pi$, a one loop part involving the $L_i$'s, and a tree level
contribution depending on some of the ${\cal O}(p^6)$ LECs $C_i$.
It should be stressed that the estimate of $f_+^{(4)}(0)$ given 
in Ref.\cite{LeutwylerRoos} is neither a two loop calculation, nor 
an estimate of the LECs that enter the two loop expression.
While the LECs giving the ${\cal O}(t)$ and the ${\cal O}(t^2)$
terms of $f_+(t)$ can in principle be obtained from the experimental
measurements of the slope $\lambda_+$ and the curvature $c_+$, 
there remain two unknown LECs in $f_+(0)$, $C_{12}$ and $C_{34}$. 
The important observation\cite{Bijnens03} here is that these same 
two LECs also appear in a combination of the scalar form factor $f_0(t)$
and of $F_K/F_\pi$. For instance,
\begin{equation}
\lambda_0 = 8\frac{M_\pi^2(M_K^2+M_\pi^2)}{F_\pi^4}(2C_{12}+C_{34})
+ \frac{M_\pi^2}{M_K^2-M_\pi^2}
\left( \frac{F_K}{F_\pi} - 1 \right) + \Delta'(0)\,,
\label{lambda0}
\end{equation}
\begin{equation}
\ c_0 = -8 \frac{M_\pi^4}{F_\pi^4}C_{12} + \Delta''(0)/2\,.
\label{c0}
\end{equation}
In the kinematical region of interest, the known function $\Delta (t)$ is well 
approximated by a polynomial\cite{Bijnens03}, $\Delta (t)=\alpha t
+\beta t^2 + \gamma t^3$. Thus, one may extract $C_{12}$ from
the knowledge of the curvature $c_0$ of $f_0(t)$, and then
get $C_{34}$ from its slope $\lambda_0$ {\it provided $F_K/F_\pi$
is known}. The reason for the emphasis\cite{Fuchs:2000hq} here 
comes from the fact that
the value usually quoted, $F_K/F_\pi = 1.22\pm 0.01$, actually 
results from the analysis of Ref.\cite{LeutwylerRoos}, and thus 
cannot be used {\it a priori}.
The effect of a variation in $F_K/F_\pi$ on $f_+(0)$ reads, 
\begin{equation}
\delta f_+(0)\vert_{F_K/F_\pi} = \frac{M_K^2 - M_\pi^2}{M_K^2 + M_\pi^2}\,
\delta\left(\frac{F_K}{F_\pi}\right)\,,
\end{equation}
and even a variation of $F_K/F_\pi$ as small as a few percents 
directly affects the value of $f_+(0)$, and thus the determination
of $\vert V_{us}\vert$, by about the same relative amount.
This assumes that all the dependence on $F_K/F_\pi$ is explicitly
shown in  Eqs. (\ref{lambda0}) and (\ref{c0}). The situation is however more
complicated, since the values of the coefficients
$\alpha,\beta,\gamma$ depend on the values of the $L_i$'s, which
are obtained from a fit\cite{Amoros01} to various input observables, {\it including the
fixed value}  $F_K/F_\pi = 1.22\pm 0.01$. A more accurate description of the
dependence on $F_K/F_\pi$ therefore requires to perform this fit for different
values of this ratio, in the range, say, from $1.17$ to $1.27$, expressing,
for instance, the numerical coefficients $\alpha,\beta,\gamma$ in the form 
$\alpha=\alpha_0 + \alpha_1(F_K/F_\pi - 1.22) + \alpha_2(F_K/F_\pi - 1.22)^2+\dots$, etc.
The situation is thus similar to the one encountered previously in a different,
but not unrelated,
context\cite{Fuchs:2000hq}, and the strategies to extract $\vert V_{us}\vert$
discussed there may be easily adapted. From Eq. (\ref{FKFpi}), one can
obtain $F_K/F_\pi$ in terms of $\vert V_{us}/V_{ud}\vert$, thus expressing
$f_+(0)$ as $1+{\cal F}(\lambda_0, c_0, \vert V_{ud}\vert, \vert V_{us}\vert)$.
Given a value of $\vert V_{ud}\vert$ and sufficiently accurate 
experimental determinations of $\lambda_0$
and of $c_0$ from the $K_{\mu 3}$ data (see the discussion in Ref.\cite{Bijnens03}
for the accuracy that is required), this would then allow to extract 
$\vert V_{us}\vert$ from the values of the $K_{\ell 3}$ branching ratios,
and then to obtain $F_K/F_\pi$ from Eq. (\ref{FKFpi}).
Independent information on $F_K/F_\pi$ can of course modify the situation.
For instance, there exists now a rather accurate determination of $F_K/F_\pi$
from partially quenched lattice data with staggered fermions\cite{MILC04}. Using
this imput allows to extract $\vert V_{us}\vert$ directly from 
Eq. (\ref{FKFpi})\cite{Marciano04}, given a value of $\vert V_{ud}\vert$. 
On the other hand, there exists also a direct, although quenched, lattice 
calculation\cite{Becirevic04} of $f_+(0)$. These new developments offer 
possibilities for cross checks. In particular, one would like to have a determination
of both $F_K/F_\pi$ and $f_+(0)$ from the same lattice simulation with
dynamical (domain wall ?) fermions, in order to check whether they satisfy
the correlation  implied by the above analysis of the two loop
ChPT expression. As far as the latter is concerned, the inclusion of
isospin breaking corrections would be welcome.

\section{Acknowledgements}
I wish to thank the organizers for the invitation to this
very informative workshop, as well as J. Bijnens and J. Gasser
for interesting discussions. Centre de Physique Th\'eorique
is Unit\'e Mixte de Recherche (UMR 6207) of CNRS, and of 
the universities Aix-Marseille I, Aix-Marseille II, 
Sud Toulon-Var, and is affiliated to the FRUMAM (FR 2291).
This work was supported in part by TMR, EC-Contract No.
HPRN-CT-2002-00311 (EURIDICE).

\end{document}